# Three-dimensional higher-order topological acoustic system with multidimensional topological states

Baizhan Xia[1,2]*, Shengjie Zheng[1,2], Liang Tong[1], Junrui Jiao[1], Guiju Duan[1], Dejie Yu[1]

1 State Key Laboratory of Advanced Design and Manufacturing for Vehicle Body, Hunan University, Changsha, Hunan, People's Republic of China, 410082

2 These authors contributed equally: B. Xia, S. Zheng.

* Correspondence to: xiabz2013@hnu.edu.cn.

**Topologically protected gapless edge/surface states are phases of quantum matter which behave as massless Dirac fermions, immunizing against disorders and continuous perturbations. Recently, a new class of topological insulators (TIs) with gapped edge states and in-gap corner states have been theoretically predicted in electric systems[1,2], and experimentally realized in two-dimensional (2D) mechanical and electromagnetic systems[3,4], electrical circuits[5], optical and sonic crystals[6-11], and elastic phononic plates[12]. Here, we elaborately design a "strong" three-dimensional (3D) topological acoustic system, by arranging acoustic meta-atoms in a simple cubic lattice. Under the direct field measurements, besides of the 2D surface propagations on all of the six surfaces, the 1D hinge propagations behaving as robust acoustic fibers along the twelve hinges and the 0D corner modes working as robust localized resonances at the eight corners are**

**experimentally confirmed. As these multidimensional topological states are activated in different frequencies and independent spaces, our works pave feasible ways for applications in the topological acoustic cavities, communications and signal-processing.**

The topological phases of matter, primarily discovered in electronic systems, have significantly renewed our understanding of condensed matter physics, and recently have extended to classical wave systems, from optical and electromagnetic systems to acoustic and elastic phononic crystals[13-18]. The key property of $d$-dimensional topological insulators (TIs) is their topologically protected ($d$-1)-dimensional gapless boundary states which immunize against disorders and continuous perturbations. Recently, a new class of TIs named as the higher-order topological insulators (HOTIs) have been theoretically proposed in quantized electronic multipole systems[1,2]. The topological phases of HOTIs are characterized by bulk polarizations but not integer topological invariants, going beyond the conversional bulk-boundary correspondence. The HOTIs with $d$-dimensions do not exhibit ($d$-1)-dimensional gapless edge states, but instead ($d$-2)-dimensional topological states on the "boundaries of boundaries". So far, the higher-order topological phases, stemming from the quantization of quadrupole moments, were experimentally realized in two-dimensional (2D) mechanical[3] and microwave systems[4] and electrical circuits[5]. Very recently, based on the skillful modulation of the inter-cell and intra-cell couplings, the higher-order topological states, expressing as gapped 1D edge states and topologically protected nontrivial in-gap zero-dimensional (0D) corner states, have been realized in conversional 2D optical and acoustic systems[6-11]. These 2D HOTIs, characterized by the bulk polarization, establish a new paradigm beyond traditional bulk-boundary physics for the design of topologically robust localized cavities.

By stacking 2D TIs in the *z* direction with appropriate interlayer couplings, the 3D "weak" TIs, which possess 2D topological surface states on some (*xz*- and *yz*- planes) but not all the surfaces, have been developed[19-21]. Recently, the 3D third-order TIs with topologically protected corner states have also been investigated[22-24]. The Su-Schrieer-Heeger (SSH) model, which is firstly known as a 1D dimerized chain, is an excellent topological model supporting topological phases in the absence of Berry curvature[25,26]. The 2D SSH models, a 2D extension of the 1D SSH chain with alternating lattice coupling in both *x* and *y* directions, have been extensively investigated in classical wave systems, including electrical circuits[27], photonic[28] and acoustic systems[29]. The topological phase of the 2D SSH model is characterized by a 2D Zak phase, and the topological edge states and even the topological corner states have been successfully predicted these 2D SSH systems. However, the experimental realization of the 3D SSH model in classical wave systems have not been reported so far. The acoustic analogues of 3D SSH model will not only provide a versatile platform for investigating the "strong" 3D TIs with 2D surfaces waves on all the surfaces, but also open perspectives for other new topological features, such as the 1D hinge states along all the hinges and the third-order topological phases expressing as the topological 0D corner states on all the corners.

Our 3D topological higher-order acoustic system is performed by starting from a discrete model which can be described by a Tight-Binding (TB) approximation. The TB approximation of the 3D SSH model is depicted in Fig. 1(a), where identical nodes are arranged in a ccD Zak phase accompanying the fractional wave polarization[30]. The real space Hamiltonian of the 3D cubic lattice is given as

$$\widehat{\mathcal{H}} = \sum_{i,j,k}[(t_x + \delta t_x x_{i,j,k})c^{\dagger}_{i+1,j,k}c_{i,j,k}$$
$$+ (t_y + \delta t_y y_{i,j,k})c^{\dagger}_{i,j+1,k}c_{i,j,k} + (t_z + \delta t_z z_{i,j,k})c^{\dagger}_{i,j,k+1}c_{i,j,k}] + H.c.$$

where $(i, j, k)$ represents a lattice point in a cubic lattice, $c^{\dagger}$ and $c$ are creation and annihilation operators at the lattice point $(i, j, k)$, $t_x$, $t_y$ and $t_z$ are transfer integrals of the cubic lattice along the $x$, $y$ and $z$ directions, respectively, and $\delta t_x$, $\delta t_y$ and $\delta t_z$ are coupling constants. $x_{i,j,k}$, $y_{i,j,k}$ and $z_{i,j,k}$ define Peierls distortions with values $x_{i,j,k} = (-1)^i$, $y_{i,j,k} = (-1)^j$ and $z_{i,j,k} = (-1)^k$. According to the parity of the lattice point $(i, j, k)$, there are two types of hopping, namely $t_w - \delta t_w$ and $t_w + \delta t_w$ ($w=x$, $y$ or $z$), representing intra-cell and inter-cell hoppings in the $x$, $y$ or $z$ direction. Due to the symmetry of our SSH model, we set $t_x=t_y=t_z=t$ and $\delta t_x=\delta t_y=\delta t_z=\delta t$, and define $\lambda = t - \delta t$ and $\gamma = t + \delta t$. The Hamiltonian matrix in the momentum space is presented in Supplementary information.

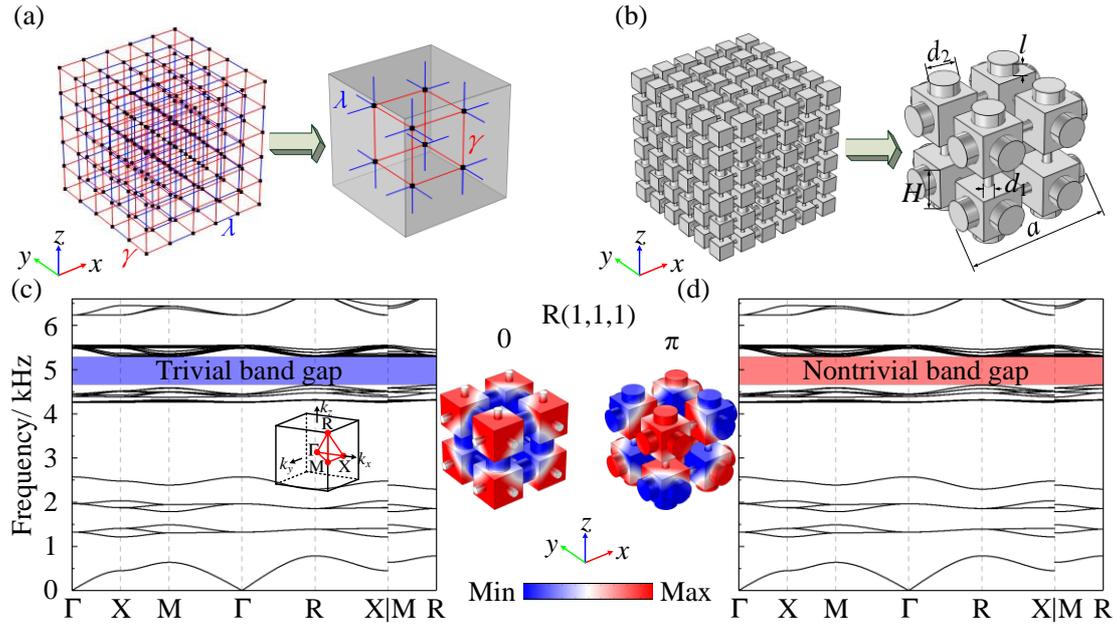

Figure 1. **(a)** Schematic presentation of the 3D SSH model. The unit cell contains eight nodes, highlighted by a gray box. The red and blue bonds represent intra-cell ($\gamma$) and inter-cell ($\lambda$) couplings. **(b)** A realistic design of the 3D acoustic system on the basis of the 3D SSH model. The insert is a unit cell including eight meta-atoms, each of which consists of a cavity and six channels. The widths of channels within and between unit cells are $d_1$ and $d_2$, representing intra-cell and inter-cell couplings. **(c)** and **(d)** Dispersion relations of the unit cell with $d_1>d_2$ and $d_1<d_2$, respectively.

The acoustic analogues of the 3D SSH model is a system consisting of acoustic meta-atoms which are cubic cavities with six channels connecting the nearest neighbours, as shown in Fig. 1(b). As the acoustic modes are strongly bounded to cubic cavities and only the nearest neighbor coupling defined by connecting channels are needed to be considered, our system is an effective realization of the 3D SSH model. Eight acoustic meta-atoms make up a unit cell, inserted in Fig. 1(b). The axis of the channel is aligned along the face-center of the cubic cavity, carefully keeping the crystalline symmetries of crystal. There are two types of channels with diameter $d_1$ and $d_2$, representing intra-cell ($\gamma$) and inter-cell ($\lambda$) hoppings between the nearest neighboring cavities. The bulk topological polarization can be achieved by varying the radio ($d_1/d_2$) of diameters of these two types of channels. The half length of the channels is $l$=10mm. The side length of the cubic cavity is $H$=35mm. The lattice constant is $a$=110mm. If the diameter of the intra-connected channel is smaller than that of the inter-connected one, namely $d_1<d_2$, it indicates a stronger resonant coupling between the adjacent unit cells. In the opposite condition, $d_1>d_2$, modes are much more strongly coupled within the unit cell.

The bulk acoustic band structures of unit cells with $d_1>d_2$ ($d_1<d_2$) along high-symmetry directions of the cubic Brillouin zone (BZ) are shown in Fig. 1c (1d). The bulk field profiles at the high-symmetric point R (inserted in Figs. 1c and 1d) clearly reveals the phase distributions of coupling between dipolar modes. It is observed that the parities of the eigen-state of the unit cell with $d_1>d_2$ exhibits an even symmetry about the *xy*-, *xz*- and *xz*-planes. Namely, the channels (with $d_1>d_2$) connecting cavities provide a zero-phase shift in the coupling between dipolar modes. However, the eigen-state of the unit cell with $d_1<d_2$ exhibits the opposite parity properties, described by the odd symmetry of modes about the *xy*-, *xz*- and *xz*-planes. It indicates that there is a $\pi$-phase shift in the coupling between dipolar modes. In this case, a $\pi$ synthetic flux is achieved by channels with $d_1<d_2$. Therefore, the unit cells with $d_1>d_2$ and $d_1<d_2$ exhibit distinct topological properties, in which the one with $d_1>d_2$ is topologically trivial and the other one with $d_1<d_2$ is topologically nontrivial.

In order to further explicitly confirm the topological phase transition between these two types of unit cells, we calculated their bulk topological polarizations characterized by an extended 3D Zak phase, seeing Supplementary information for detail. When $\gamma>\lambda$ which is defined by $d_1>d_2$, it is yielded that $v_x = 0$ for all $(k_y, k_z)$, $v_y = 0$ for all $(k_x, k_z)$ and $v_z = 0$ for all $(k_x, k_y)$, leading to a trivial 3D bulk polarization with $\mathbf{P} = (p_x, p_y, p_z) = (0,0,0)$. In contrast, when $\gamma<\lambda$ which is defined by $d_1<d_2$, a topologically nontrivial 3D bulk polarization with $\mathbf{P} = (p_x, p_y, p_z) = (1/2, 1/2, 1/2)$ is yielded. Thus, our topological system is different from previous TIs, as its topological phase is characterized by a non-zero 3D Zak phase $\mathbf{\theta} = (\pi, \pi, \pi)$, accompanying with the fractional wave polarization $\mathbf{P} = (1/2, 1/2, 1/2)$.

The intrinsic characteristic of TIs is the emergence of surface/hinge states which are localized at surfaces and hinges of topological nontrivial systems. To simulate these topological features, we construct a supercell consisting of $3 \times 3$ unit cells aligned in the *xy*- plane with periodic boundary condition along the *z*-direction, as shown in Fig. 2(a). When $d_1>d_2$, there is no surface or hinge state in the complete band gap (seeing Fig. 2b), indicating that this supercell is trivial which matches with the theoretical prediction. On the contrary, when $d_1<d_2$, the surface and hinge states can be clearly observed in the complete band gap (seeing Fig. 2c), which indicates that the bulk polarization of this supercell is topologically nontrivial. Fig. 2d is the field profile which corresponds the surface band marked by blue in Fig. 2c. It shows that the acoustic energy is well localized at the surfaces of supercell, except of hinges. For the hinge states marked by green in Fig. 2c, it can be observed from Fig. 2e that the acoustic energy is well concentrated along the hinges of supercell, but keeping the bulk and surfaces insulating.

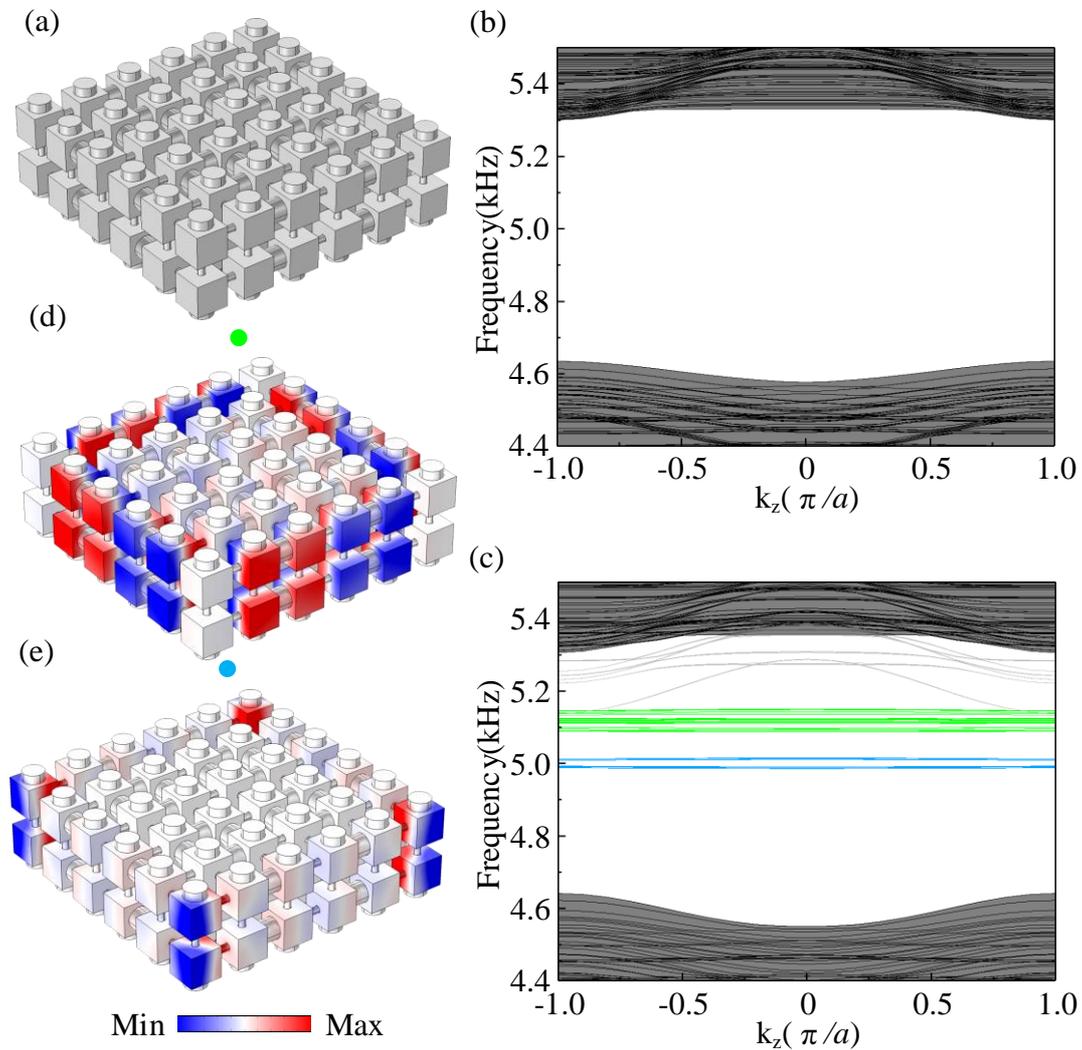

Figure 2. **(a)** A sketch of the supercell composed of $3 \times 3$ unit cells aligned in the x-y plane with periodic boundary condition along the z-direction. **(b)** Simulated dispersions of the supercell composed of $3 \times 3$ trivial unit cells with $d_1 > d_2$. In the bandgap of bulk states, there is no any surface and hinge states. **(c)** Simulated dispersions of the supercell composed of $3 \times 3$ nontrivial unit cells with $d_1 < d_2$. The surface (blue) and hinge (green) states emerge in the bandgap of bulk states. (d) and (e) The simulated displacement field profiles of surface and hinge states, respectively.

To clearly observe the surface and hinge modes and verify the emergence of the corner mode, we design two acoustic networks consisting of $3 \times 3 \times 3$ unit cells (as depicted in Fig. 3a and 3b and seeing simulation of method for details) whose eigenfrequency are calculated by the first-principle finite element method. The first acoustic network consisting of trivial unit cells with $d_1>d_2$ is a trivial sample. Numerically calculated eigenfrequency of this acoustic network, as presented in Fig. 3c, shows that there are no surface, hinge and corner modes in the complete bandgap between the lower- and the higher-frequency bulk modes. On the contrary, the second acoustic network is a topologically protected one consisting of nontrivial unit cells with $d_1<d_2$. The surface, hinge and corner modes emerge in the topologically nontrivial bandgap between the lower- and the higher-frequency bulk modes. Simulated field profiles of these bulk, surface, hinge and corner states are presented in Figs. 3e, 3f, 3g and 3h, respectively. For the bulk mode, the acoustic energy is distributed in the bulk of acoustic network. To clearly observe the bulk state, the fifteen surface cavities are hidden in Fig. 3e. For the surface mode, the acoustic energy is distributed on the six surfaces of acoustic network, except of the twelve hinges (as shown in Fig. 3f), which is different from the traditional 3D first-order TIs with surface states on the whole surfaces including hinges[19-21]. For the hinge mode, the acoustic energy is strongly localized along the twelve hinges of acoustic network, except of the eight corners, as presented in Fig. 3g. For the corner mode which is the essential feature of the third-order topological state, the acoustic energy is extremely concentrated at the eight corners of acoustic network, decaying rapidly along the hinges, on the surfaces and in the bulk, as shown in Fig. 3h. These simulated pressure field distributions show that the surface, hinge and corner states are spatially separated from each other. Namely, the hinge mode cannot be activated by the surface state and the corner mode cannot be activated by the hinge state. Furthermore, due to the nontrivial bulk polarization of this acoustic system, the surface, hinge and corner states exhibit a good immunity against defects, which can be confirmed by several imperfect acoustic networks deliberately designed in the Fig. S1 of Supplementary information.

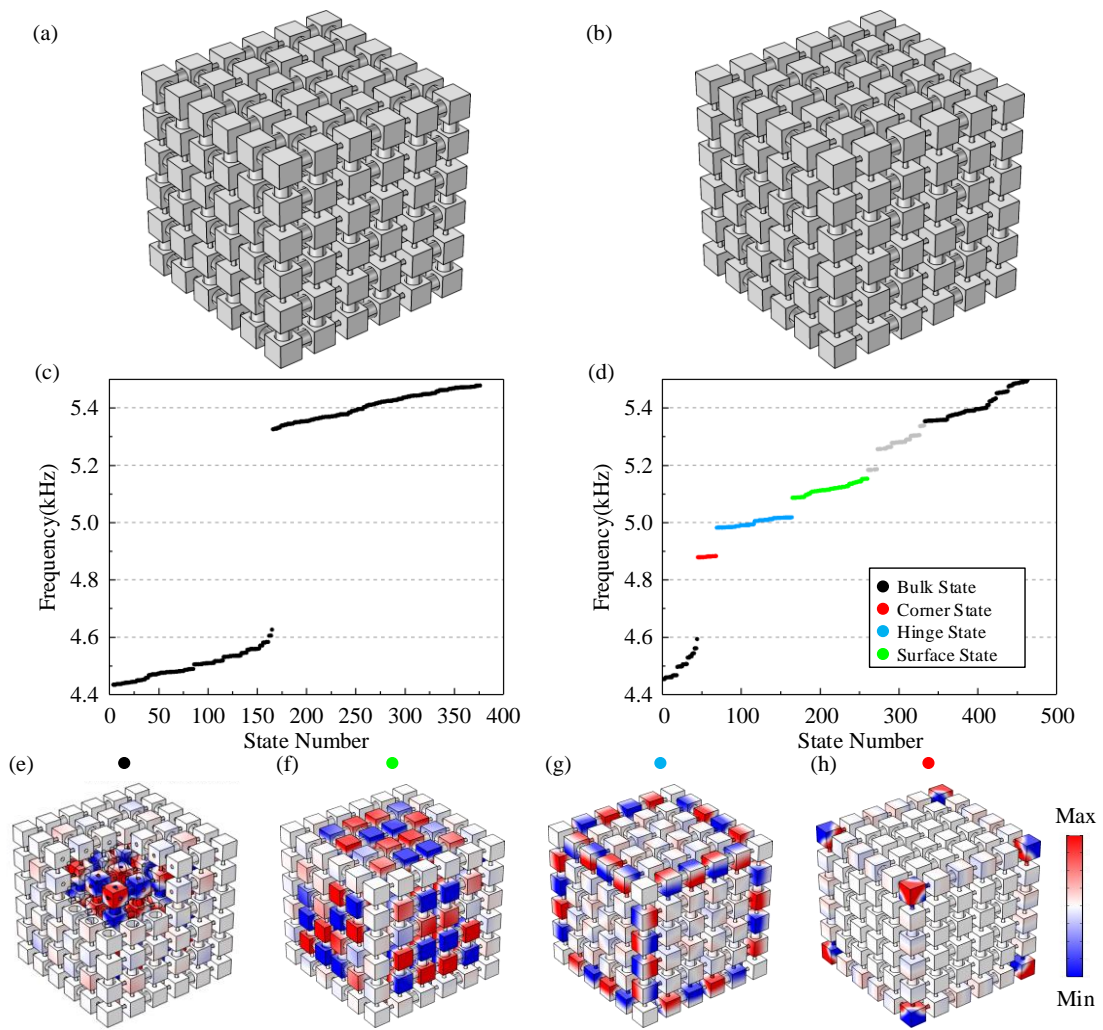

**Figure 3.** **(a)** and **(b)** Numerical models of the higher-order acoustic networks with $3 \times 3 \times 3$ trivial and nontrivial unit cells, respectively. **(c)** and **(d)** Numerically evaluated eigenfrequencies

of acoustic networks with trivial and nontrivial unit cells. Black, green, blue and red dots denote the bulk, surface, hinge and corner modes, respectively. **(e)-(h)**, Simulated displacement field profiles of bulk (4466.1 Hz), surface (5135.3 Hz), hinge (5011.4Hz) and corner (4883.1 Hz) modes.

To experimentally visualize these surface, hinge and corner states, we fabricate a topologically nontrivial acoustic sample by the 3D printing technique (seeing Fig. 4a and seeing methods for details). The surface, hinge and corner transmission spectra are presented in Fig. 4b. To measure the surface transmission spectrum (the green curve), we place the source near a cavity on the bottom surface and detect the response of the cavity on the top surface. We clearly observe a high peak around 5136Hz, being consistent with the numerically predicted surface mode. To measure the hinge transmission spectrum (the blue curve), we place the source near the cavity at the left hinge of the bottom surface and detect the response of the cavity at the left hinge of the top surface. There is a high peak emerging around 5010Hz which is also in record with the numerically predicted hinge mode. Last, to measure the corner transmission spectrum (the red curve), the source is placed near the left-back corner of the bottom surface and detect the response of the left-back corner of the top surface. It yields a high peak emerges around 4888Hz which experimentally confirms the third-order topological feature of our 3D acoustic system. The bulk transmission spectrum with a low efficiency in the complete band gap is also plotted in Fig. 4b. Frequency response spectra of other surfaces, hinges and corners are shown in Figs. S2a, S2b and S2c of Supplementary information, respectively, providing further evidence of the higher-order topological phases of our acoustic system.

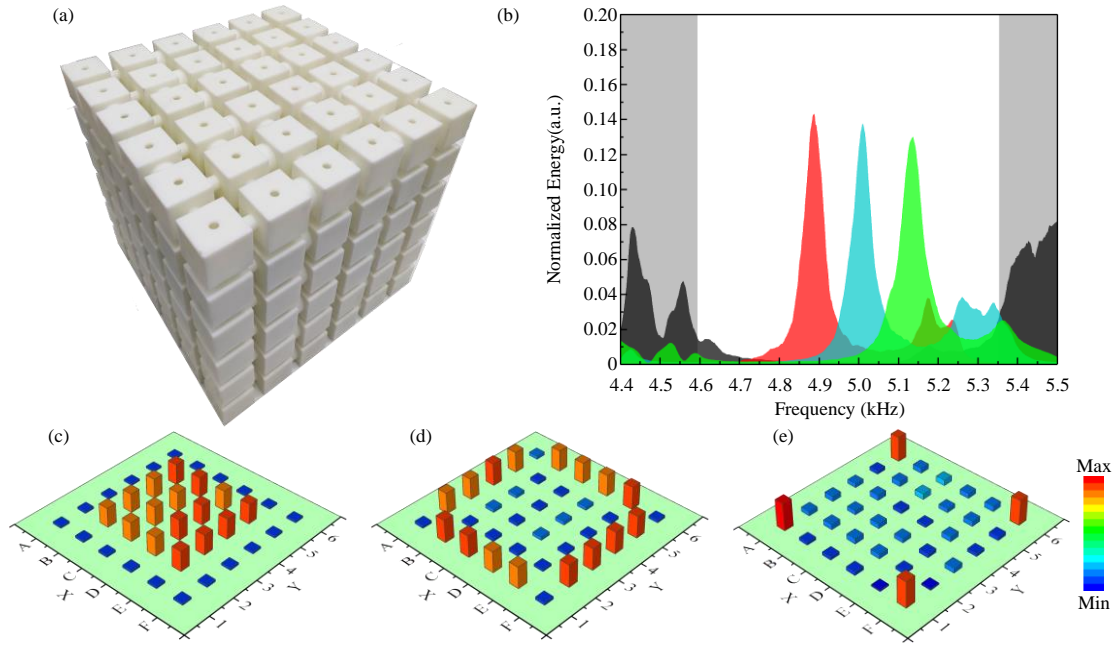

Fig. 4 | **(a)** Photograph of the fabricated acoustic system consisting of $3 \times 3 \times 3$ nontrivial unit cells. **(b)** Measured bulk (black), surface (blue), hinge (green) and corner (red) transmission spectra for a topological hexagon-shaped sample. **(c)**, **(d)** and **(e)** Measured pressure field profiles for the surface (5136Hz), hinge (5010Hz) and corner (4888Hz) states, respectively. Full wavefield numerical simulations of the topological acoustic system are constructed in the Fig. S1 of Supplementary information.

Three measured pressure field profiles of the surface (5136Hz), hinge (5010Hz) and corner (4888Hz) states are plotted in Figs. 4c, 4d and 4e, respectively. To explicitly excite the surface, hinge and corner states, the sources are placed at the cavity on the bottom surface, the cavity at the left hinge of the bottom surface and the left-back corner of the bottom surface, respectively. The measurements are done by scanning the acoustic pressure amplitudes of all cavities on the top surface (seeing method for details). At the frequency of 5136Hz, the measured pressure field profile shows that the acoustic energy mainly distributes on the surface, except of the four hinges. Namely, the hinge response is suppressed under the surface state. At the frequency of 5010Hz, the acoustic energy is localized along the four hinges, rapidly decaying away from the four hinges. It should be noted that the response of four corners are considerably weaker when compared with the hinge response, indicating that the corner response is suppressed under the hinge state. At the frequency of 4888Hz, we observe a typical behavior of the 0D corner mode. Namely, the acoustic energy almost concentrates at the four corners, rapidly decaying along hinges and on surfaces. This is an overwhelming experimental evidence of the third-order topology of our 3D acoustic system.

In conclusion, we realized a 3D topological higher-order acoustic system in a cubic lattice with a nontrivial 3D bulk polarization. Multidimensional topological states of our acoustic system, including the 2D surface states, the 1D hinge states and the 0D corner states, were observed in measured transmission spectra and measured pressure field profiles. These multidimensional topological states were unambiguously separated from each other in both the frequency spectra and the space distributions, offering a flexible and well-controlled platform for the unprecedented wave modulation, from the 2D surface propagations, the 1D hinge trappings to the 0D localized resonances. We also anticipate that our 3D higher-order acoustic system may inspire more attentions on the development of other 3D higher-order topological integrated chips, from photonic, electrical, thermal to mechanical systems.

We note a recent work, which realize the acoustic analogue of a quantized octupole topological insulator (QOTIs) in 3D acoustic systems, by using negatively coupled resonators[31].

## Acknowledgements

This work is supported by the Innovative Research Groups of the National Natural Science Foundation of China (Grant No. 51621004).

## Author contributions

B.X. and S.Z initiated the study and guided the work. B.X. and S.Z designed the experiment. B.X., S.Z. and L.T. made the simulations. B.X., S.Z., J.J. and G.D. fabricated samples, carried out the measurement and analyzed data. B.X., S.Z and D.Y. provided the theoretical explanations. B.X., S.Z and D.Y. wrote the manuscript, with input from all co-authors. B.X. supervised the project.

## Competing interests

The authors declare no competing interests.

## Methods

**Simulations**

The acoustic system consisting of air cavities connected by six channels are made of photosensitive resin 9400 SLA (modulus 2,765 MPa, density 1.3 g/cm$^3$) which can be considered to be rigid. The geometries of the unit cell in Fig. 1b are as follow: the lattice constant of the simple cubic crystal is $a$=110mm, the side length of the cavity is $H$=35mm, the half length of the channel is $l$=10mm, the diameters of the intra-cell channels are $d_1$=24mm for the trivial unit cell and $d_1$=8mm for the nontrivial unit cell, the diameters of the inter-cell channels are $d_1$=8mm for the trivial unit cell and $d_2$=24mm for the nontrivial unit cell. The cavities are coupled through cylindrical channels and the coupling strength of the modes is adjusted by changing the diameter of channels. For the trivial unit cell, the full-wave simulations are accurately carried by a commercial finite element analysis software (COMSOL Multiphysics). The mass density and the sound velocity in air are 1.21 kg/m$^3$ and 343 m/s, respectively. For the bulk band structures and modes of unit cells (in Fig. 1c-1d), the Floquet periodic boundary conditions are imposed on all the six surfaces to form an infinite cubic lattice. For bulk and edge band structures and modes of supercells (in Fig. 2b and 2c), the rigid boundary conditions are imposed on the left, right, front and back surfaces, while the Floquet periodic boundary conditions are imposed on the top and bottom surfaces of supercells. The acoustic pressure fields are used to describe the surface and hinge states of supercell (Fig. 2d and 2e). In numerical simulations of Fig. 3, we used a large cubic sample consisting of $3 \times 3 \times 3$ unit cells for a clear separation of corner, hinge and surface modes. The simulated field profiles of the cubic samples were performed by applying the rigid boundary conditions on the six surfaces of samples. The "state number" is the "eigenfrequency number". For Fig. 3c and 3d, 376 and 463 eigenfrequencies in the vicinity of the bandgap are considered. These eigenfrequencies are calculated by solving the eigenvalue equations of the cubic sample. The acoustic pressure fields are used to describe the eigenstates of bulk, surface, hinge and corner modes of samples (Fig. 3e-3h). In the simulations of excitations (Fig. S1), all the six surfaces are set as be rigid and the locations of sound sources are marked by Yellow horns. In Figs. S1d-S1f, a cavity is removed from the surface, hinge and corner of the sample, marked by the green circle. In Figs. S1g-S1i, two cavities are removed from the surfaces, hinges and corners of the sample, marked by green circles.

**Experiments**

The measured cubic sample in Fig. 4a is fabricated by the 3D printing technique, using the photosensitive resin 9400 SLA (modulus 2,765 MPa, density 1.3 g/cm$^3$). The sample is made of the nontrivial unit cells, in which the external diameters of intra-cell and inter-cell channels are set to be $d_{1e}$=14mm and $d_{2e}$=30mm, respectively. The thick wall of the cavities and channels (with $t$=3mm) can be considered to be hard in the experimental test. The machining tolerance is about 0.1mm. The left, right, front and back surfaces of sample are hermetically sealed, while the small probe holes with the diameter $d_0$=8.00mm were intentionally introduced on the top and bottom surfaces. Each cavity has a probe hole. In the experimental process, only the probe holes excited by the sound source and measured by the microphone are open, while the other probe holes are blocked by the plexiglass plate with thickness $t_0$=10mm to suppressing the excessive loss. Experiments are conducted by a horn through a funnel-shaped tube whose long mouth can inserted into the bulk of sample. In Figs. 4b, 4c and S2a, the funnel-shaped tube inserts into the cavity on the bottom surface to excite the surface states. In Figs. 4b, 4d and S2b, the funnel-shaped tube inserts into the cavity at the left hinge of the bottom surface to excite the hinge state. In Figs. 4b, 4e and S2c, the funnel-shaped tube inserts into the cavity of the left-back corner of the bottom surface. The sound source input is a white noise signal. The sound signal is measured by a movable microphone (BK Type 4944-A) and analyzed in LMS SCADAS III. The sampling range is 10Hz-6400 Hz with an increment of 1 Hz. For the transmission spectra measurements of surface, hinge and corner response in Fig. 4b, the movable microphone inserts into the cavity on the bottom surface, the cavity at the left hinge of the bottom surface and the left-back corner of the bottom surface, respectively. For the transmission spectra measurements of two surfaces in Fig. S2a, the movable microphone inserts into the cavities on the bottom and top surfaces. For the transmission spectra measurements of eight hinges in Fig. S2b, the movable microphone inserts into the cavities of seven hinges on the bottom and top surfaces, except of the left one on the bottom surface. For the transmission spectra measurements of seven corners in Fig. S2c, the movable microphone inserts into the other seven corners, except of the left-back one of the bottom surface. During three measured pressure field profiles in Figs. 4c-4e, the movable microphone scans all probe holes on the top surface.

## Note 1. Hamiltonian matrix in the momentum space

Under a Fourier transformation, the Hamiltonian matrix in the momentum space is expressed as

$$\mathcal{H}(\kappa) = \begin{pmatrix} 0 & \mathcal{M}(\kappa) \\ \mathcal{M}^\dagger(\kappa) & 0 \end{pmatrix}$$

$$\mathcal{M}(\kappa) = \begin{pmatrix} h_{11} & h_{12} & h_{13} & 0 \\ h_{21} & h_{22} & 0 & h_{24} \\ h_{31} & 0 & h_{33} & h_{34} \\ 0 & h_{42} & h_{43} & h_{44} \end{pmatrix}$$

where $h_{11} = h_{44} = \gamma + \lambda exp(-ik_x)$ and $h_{22} = h_{33} = h_{11}^*$, $h_{12} = h_{34} = \gamma + \lambda exp(-ik_y)$ and $h_{21} = h_{43} = h_{12}^*$, and $h_{13} = h_{24} = \gamma + \lambda exp(-ik_z)$ and $h_{31} = h_{42} = h_{13}^*$. This Hamiltonian model satisfies the time-reversal symmetry $\mathcal{H}^*(\kappa) = \mathcal{H}(-\kappa)$, and the chiral symmetry $\tau_3 \mathcal{H}(\kappa) \tau_3 = -\mathcal{H}(\kappa)$ as it is block-diagonal. Here, $\tau_3$ is the third Pauli matrix.

## Note 2. The bulk topological polarization.

The bulk topological polarization characterized by an extended 3D Zak phase is given by the following expression

$$\mathbf{P} = \frac{1}{2\pi} \int dk_x dk_y dk_z \text{Tr}[\mathcal{A}(k_x, k_y, k_z)]$$

where $\mathcal{A} = \langle \psi | i \partial_\mathbf{k} | \psi \rangle$ is the Berry connection and the integration over the first BZ. The $x$, $y$ and $z$ component of the bulk polarization can be straightforwardly rewritten as $p_x = \frac{1}{2\pi} \int dk_y dk_z v_x(k_y, k_z)$ , $p_y = \frac{1}{2\pi} \int dk_x dk_z v_y(k_x, k_z)$ and $p_z = \frac{1}{2\pi} \int dk_x dk_y v_z(k_x, k_y)$, respectively.

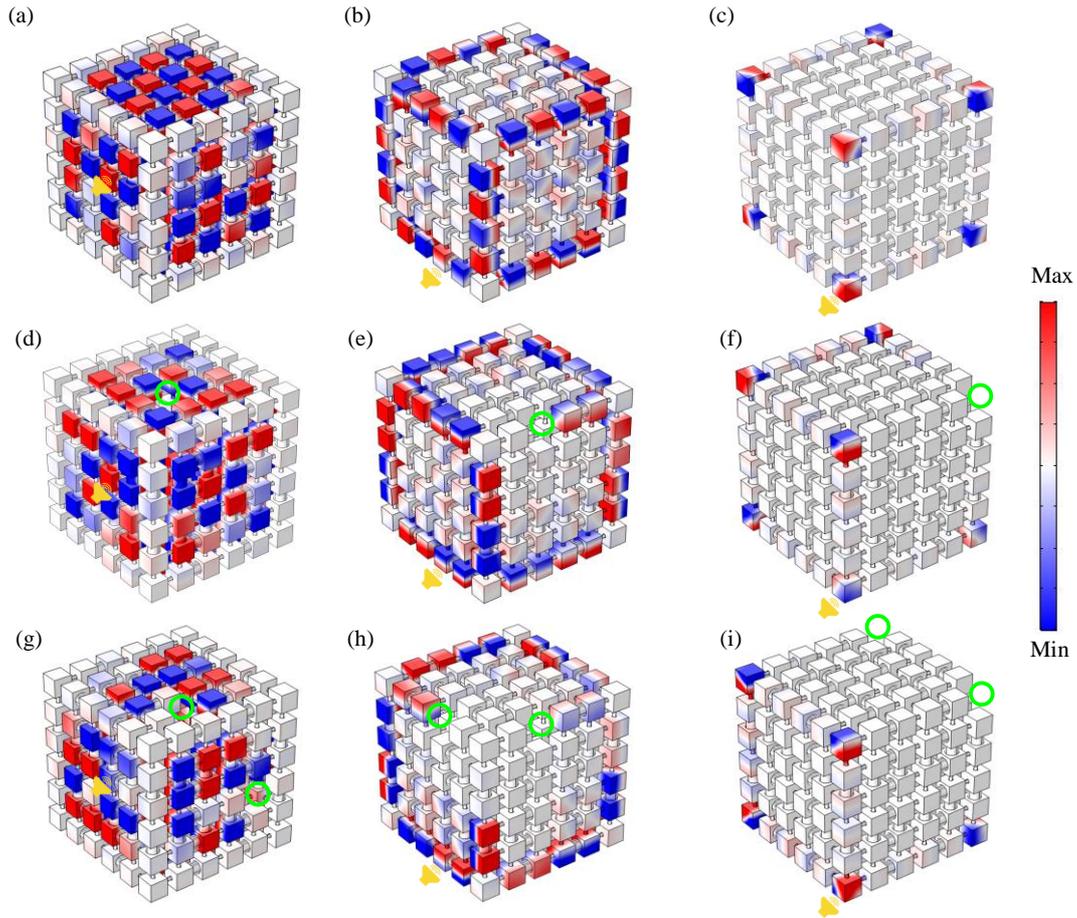

Fig. S1. The full wavefield numerical simulations of surface, hinge and corner modes in the cubic samples. (**a**)-(**c**) The simulations of the perfect cubic samples. (**d**)-(**f**) The simulations of the cubic samples with a defect. (**d**) A cavity on the top surface is removed. (**e**) A cavity on the front hinge of the top surface is removed. (**f**) A cavity on the right-front corner of the top surface is removed. (**g**)-(**i**) The simulations of the cubic samples with two defects. (**g**) Two cavities on the top and front surfaces are removed. (**e**) Two cavities on the front and left hinges of the top surface are removed. (**f**) Two cavities on the right-front and right-back corners of the top surface are removed. The removed cavities are marked by green circles. The locations of sound sources are marked by Yellow horns.

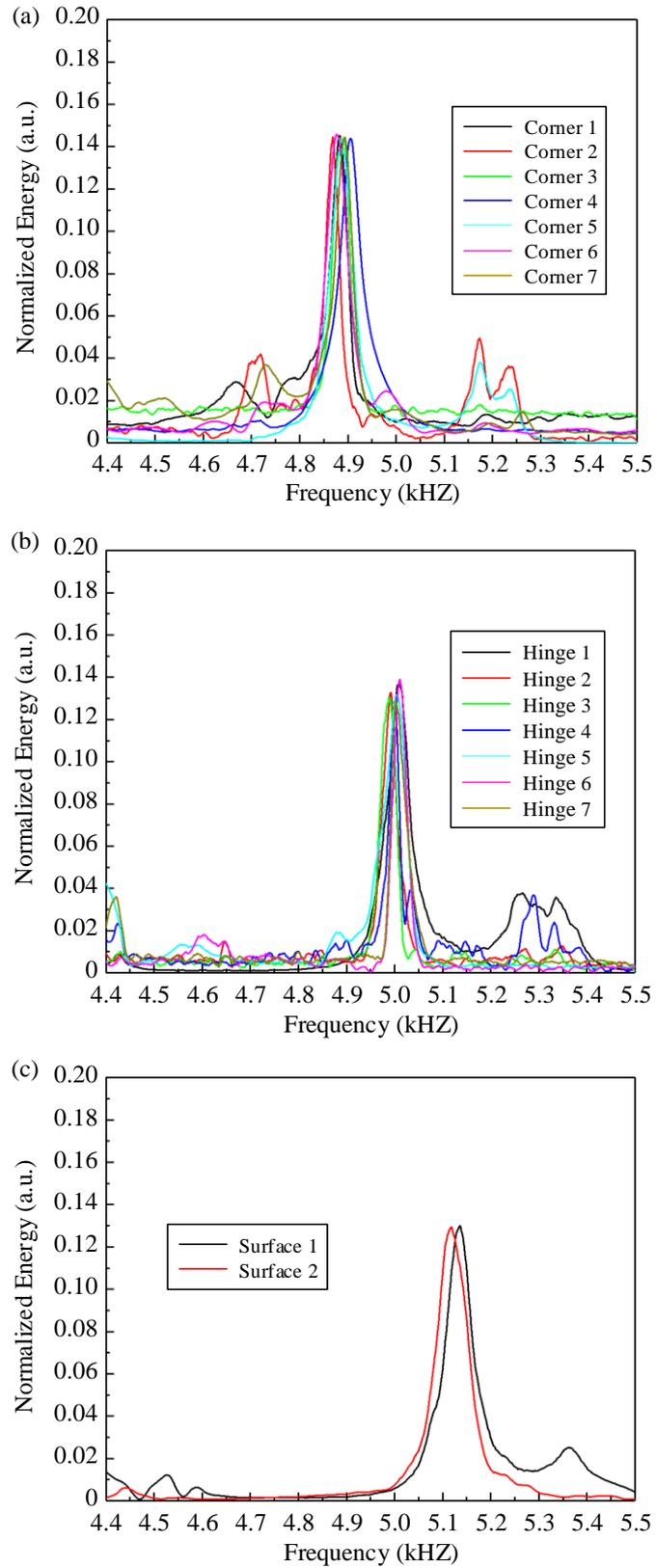

Fig. S2. Frequency response spectra of two surfaces, seven hinges and seven corners. Resonant peaks of these frequency response spectra are clearly observed near the surface, hinge and corner modes, respectively. Little fluctuations of resonant peaks of surface, hinge and corner responses suffer from the fabricating errors of the 3D printing. The fluctuations of resonant frequencies of meta-atoms will lead to the chiral symmetry reduction of crystal.